# Toward Total Recall: Enhancing Data FAIRness through AI-Driven Metadata Standardization

Sowmya S. Sundaram, Rafael S. Gonçalves, and Mark A. Musen

Stanford Center for Biomedical Informatics Research, Stanford University, Stanford, California, USA

*Corresponding author(s). E-mail(s): sowmyasm@stanford.edu; Contributing authors: goncalves@stanford.edu; musen@stanford.edu

**Abstract:**

Scientific metadata often suffer from incompleteness, inconsistency, and formatting errors, which hinder effective discovery and reuse of the associated datasets. We present a method that combines GPT-4 with structured metadata templates from the CEDAR knowledge base to automatically standardize metadata and to ensure compliance with established standards. A CEDAR template specifies the expected fields of a metadata submission and their permissible values. Our standardization process involves using CEDAR templates to guide GPT-4 in accurately correcting and refining metadata entries in bulk, resulting in significant improvements in metadata retrieval performance, especially in *recall*—the proportion of relevant datasets retrieved from the total relevant datasets available. Using the BioSample and GEO repositories maintained by the National Center for Biotechnology Information (NCBI), we demonstrate that retrieval of datasets whose metadata are altered by GPT-4 when provided with CEDAR templates (GPT-4+CEDAR) is substantially better than retrieval of datasets whose metadata are in their original state and that of datasets whose metadata are altered using GPT-4 with only data-dictionary guidance (GPT-4+DD). The average recall increases dramatically, from 17.65% with baseline raw metadata to 62.87% with GPT-4+CEDAR. Furthermore, we evaluate the robustness of our approach by comparing GPT-4 against other large language models, including LLaMA-3 and MedLLaMA2, demonstrating consistent performance advantages for GPT-4+CEDAR. These results underscore the transformative potential of combining advanced language models with symbolic models of standardized metadata structures for more effective and reliable data retrieval, thus accelerating scientific discoveries and data-driven research.

# Introduction

Effective data sharing can be improved by adhering to the FAIR guiding principles [1], ensuring that data are Findable, Accessible, Interoperable, and Reusable. A key requirement for making data FAIR is high quality metadata [2], which provide information about their associated data such as their origin, purpose, usage, and the conditions under which the data were collected. These metadata play a pivotal role in facilitating the reproducibility and organization of the data and in enhancing their discoverability. Nevertheless, metadata in scientific repositories are often incomplete, inconsistent, and incorrectly formatted, hindering data discoverability and reuse [3 ,4]. Standardizing metadata is therefore a necessary process for ensuring that datasets are properly described and accessible, enabling their reuse, integration, and secondary analysis. However, manually improving metadata quality is a complex and time-consuming process due to variability in community standards and subtleties in applying such standards. In this paper, we introduce an automated approach to standardizing metadata by leveraging structured metadata templates—machine-readable specifications that define required metadata fields and permissible values—and large language models (LLMs) such as GPT-4 [5]. We evaluate the effectiveness of our approach through key information-retrieval metrics, primarily focusing on *recall*, which measures the proportion of relevant metadata records correctly retrieved out of all relevant metadata records. Additionally, we assess *precision*, the proportion of retrieved metadata records that are relevant, and the F1-score, a balanced measure combining precision and recall. Our study demonstrates that AI-driven metadata standardization significantly enhances these retrieval metrics, greatly improving the discoverability and usability of scientific data.

Scientific metadata consist of pairs of field names and field values that describe datasets derived from experimental results. In Figure 1, we show an example metadata record taken from the BioSample repository of the National Center for Biotechnology Information (NCBI) [6]—a public repository that stores metadata about biological samples used in genomic, transcriptomic, and other biomedical studies. In this example, the field name is given as "tissue" and the field value is given as "lung cancer." Even without knowing the details of the original sample, we know that the metadata field value is inaccurate, as *lung cancer* is not a type of tissue. Searches for scientific datasets primarily involve querying metadata. Consequently, a researcher querying for appropriate tissue values would overlook this record and potentially other useful records during their search.

**Fig. 1** Metadata record from BioSample, where the black box highlights a field name–field value pair. In this example, the field name "tissue" is wrongly associated with the field value "lung cancer." Orange boxes mask identifying information.

A key strategy for improving the quality of scientific metadata involves the development of discipline-specific metadata standards that draw upon both reporting guidelines (that provide standard field names) and ontologies (that provide standard field values). For example, in the functional genomics community, the Minimum Information About a Microarray Experiment (MIAME) guideline specifies essential metadata attributes needed for reproducibility and reusability of microarray data, such as details about the sample that was studied, the experimental design, data processing, and normalization methods. By establishing clear, structured reporting standards, MIAME improved the consistency, comparability, and overall utility of microarray datasets, facilitating more effective data sharing and secondary analysis within the genomics community [7]. Ontologies provide structured, standardized vocabularies that facilitate consistent annotation across datasets, supplying systematized values for the fields enumerated by reporting guidelines. For example, UBERON [8] is a multi-species anatomy ontology that offers a standardized vocabulary for anatomical structures in animals and is increasingly used to annotate tissue-related datasets. For metadata standards to be effective, however, they must be tailored to specific research domains, as different scientific disciplines require metadata specifications that reflect their unique experimental situations. A promising approach is for research communities to define discipline-specific metadata reporting guidelines and to render those guidelines as structured templates. The Center for Expanded Data Annotation and Retrieval (CEDAR) Workbench [9] offers scientists a collaborative and efficient means to create machine-actionable, structured metadata templates for authoring standards-adherent metadata. A CEDAR template (Figure 2) is a machine-readable specification of the expected metadata fields and their permissible values, which may be sourced from multiple ontologies or controlled vocabularies available in the BioPortal repository [10]—the most comprehensive, open archive of biomedical ontologies. By encoding well-established reporting guidelines using CEDAR templates, researchers can ensure that metadata are standardized and machine-actionable, ultimately enhancing data discovery and secondary analysis. Recent work by our laboratory [11, 12] has demonstrated that encoding community metadata standards as structured CEDAR templates provides a scalable, automated way to enhance the FAIRness of scientific data by improving metadata quality and standardization.

**Fig. 2**: A screen capture depicting the CEDAR metadata entry form derived from the template for BioSample. Field names specified in the data dictionary, such as "tissue" and "organism," are listed in the template. CEDAR recommends options for the "tissue" field from a branch of the Uberon ontology.

Automated metadata cleaning inherently involves natural language processing, a challenge for which large language models (LLMs) have recently proven particularly effective. Contemporary research efforts [13, 14] have explored the use of LLMs to improve metadata quality, including efforts to curate specific fields such as cell line names in BioSample. These studies show that targeted metadata curation with LLMs can substantially improve the quality and usability of key metadata fields. However, the focus of these studies has largely been limited to a small number of fields—such as cell line names—within otherwise well-structured metadata. In contrast, real-world metadata authoring often requires reasoning over entire reporting guidelines with numerous interdependent and heterogeneous fields, presenting additional challenges in scalability and generalizability that prior work has not addressed. This limitation underscores the need for approaches that can operate in more complex metadata environments where multiple interdependent fields must be curated in concert. Our recent work [15] demonstrates that automated multi-field metadata curation is a challenging task for LLMs that may be addressed by augmenting LLMs with knowledge derived from expert feedback. In this paper, we present the quantifiable effects of using structured metadata templates in CEDAR in tandem with LLMs to automatically correct legacy metadata. We evaluate our method on a large dataset of 4,800 samples drawn half from BioSample and half from the Gene Expression Omnibus (GEO) [16], measuring the impact of metadata improvements on search outcomes. The Gene Expression Omnibus (GEO) is a public database at the NCBI that archives and freely distributes high-throughput gene expression and other functional genomics data submitted by the scientific community. We also verify whether the benefits of using metadata templates to guide metadata standardization are consistent across different LLMs, including GPT-4, Large Language Model Meta AI (LLaMA-3) [17], and MedLLaMA-2 [18]. While the FAIR principles are widely endorsed for their theoretical benefits, a lack of quantitative analysis has made it difficult to assess their practical effects on dataset searchability. A key contribution of our work is the quantification of metadata retrieval improvements following metadata standardization.

**Materials and methods**

Our work aims to enhance the quality of scientific metadata by leveraging LLMs and the CEDAR structured metadata knowledge base. Specifically, we evaluate how improvements in metadata quality affect search-related metrics, with a focus on recall performance. We begin by providing a detailed description of the datasets used in our study, including the scope, structure, and selection criteria of the metadata analyzed.

**Dataset**

For our experiment, we used metadata records from two distinct databases: BioSample and GEO. We developed queries (shown in Table 1) to retrieve metadata records from both BioSample and GEO related to three types of cancer: lung cancer, liver cancer, and ovarian cancer.

| Disease | Database | Query |
|---|---|---|
| Lung cancer | BioSample | lung cancer[All Fields] AND "human 1 0"[filter] |
| | GEO | lung cancer[All Fields] AND human[Organism] |
| Liver cancer | BioSample | liver cancer[All Fields] AND "human 1 0"[filter] |
| | GEO | liver cancer[All Fields] AND human[Organism] |
| Ovarian cancer | BioSample | ovarian cancer[All Fields] AND "human 1 0"[filter] |
| | GEO | ovarian cancer[All Fields] AND human[Organism] |

**Table 1** Queries used to retrieve metadata records from the BioSample and GEO databases.

For each of the three types of cancer, we extracted a sample of 800 randomly selected records from each dataset. We initially sampled 1,000 records for each query, removed those with XML formatting errors, and selected the maximum uniform number of well-formatted records across all queries, which was 800. Our test corpus thus comprises 4,800 records: 2,400 from BioSample and 2,400 from GEO.

**Approach**

To automate the correction of metadata records, our first method was to instruct an LLM to use the BioSample data dictionary [19]—which defines allowed metadata field name–field value pairs, formats, and descriptions—to correct the metadata record. This process and data dictionary is illustrated in Figure 3. The GEO repository lacks a similarly detailed data dictionary. Although GEO recommends adherence to the widely adopted MIAME guidelines, these guidelines primarily outline broad metadata reporting requirements—specifying the types of information to include—rather than offering structured descriptions of individual field–value pairs. Consequently, for metadata fields specific to GEO, we reused relevant portions of the BioSample data dictionary.

We then designed a second method that uses structured metadata templates built with CEDAR as guiding mechanisms for LLMs to correct metadata. CEDAR templates contain machine-readable restrictions on fields that can assist an LLM in determining when a value is inappropriate for a field. These restrictions include data-type restrictions (e.g., string, date, integer), value-set restrictions, and ontological restrictions that ensure that metadata entries are consistent with established standards. Given its comprehensive repository of metadata templates, CEDAR is a natural choice for supporting metadata-related endeavors. For example, as shown in Figure 2, CEDAR has in its library a BioSample metadata template where the "tissue" field only takes values from the Uberon ontology—a comprehensive multi-species anatomy ontology—thus restricting the user to select standard terms. A dedicated CEDAR template for GEO was not available, so resused relevant sections of the BioSample CEDAR template.

We evaluated the described framework using different LLMs (GPT-4, LLaMA-3 and MedLLaMA2) and our test dataset. In our test data, we used two versions of each metadata collection—one that is augmented based on a data dictionary (LLM+DD), and another that is augmented based on information in a CEDAR template (LLM+CEDAR). We searched the standardized metadata collections using four representative queries targeting the metadata field "tissue." In these queries, we adopted the syntax *field:value*, indicating the metadata field followed by its desired value. Specifically, our queries were *tissue:lung*, *tissue:blood*, *tissue:liver* and *tissue:ovary,* adjusted for the relevant cohort. The search was based on exact string matching between the standardized metadata field values and the query terms. Specifically, a metadata record was counted as retrieved if the value in its "tissue" field matched the query string exactly, without partial matching, stemming, or synonym expansion. We then measured commonly used search metrics, including precision, recall and F1-score. The F1-score is the harmonic mean of precision and recall, providing a single summary metric that balances both measures.

**Input to GPT-4: Metadata Record and Data Dictionary**

Convert the record: "age:67 — sex:female — tissue:lung cancer" to the format given by the BioSample data dictionary:

"Name — Description — Value format
age — age at the time of sampling; relevant scale depends on species and study, e.g. could be seconds for amoebae or centuries for trees — {float}{unit}
...
tissue — Type of tissue the sample was taken from — {term}
disease — list of diseases diagnosed; can include multiple diagnoses. the value of the field depends on host; for humans the terms should be chosen from DO (Disease Ontology), free text for non-human. — {term}"

**Output:**
biosample provider: NA
age: 67
sex: female
disease: lung cancer

**Fig. 3** GPT-4 Correction of BioSample metadata using a data dictionary. The prompt given to GPT-4 includes both the metadata to be corrected and the entire data dictionary to which BioSample records should adhere.

**Input to GPT-4: Metadata Record and CEDAR template**

Convert the record: "age:67 — sex:female — tissue:lung cancer" to the format given by the CEDAR template

"Name — Description — Comments
age — age at the time of sampling; relevant scale depends on species and study, e.g. could be seconds for amoebae or centuries for trees — {float}{unit}
tissue — type of tissue sample — **Must be from Uberon ontology**
...
disease — Name of the disease — **Must be from Disease Ontology (DO) ontology**"

**Output:**
biosample accession: NA
organism: Homo sapiens
age: 67
sex: female tissue: lung
disease: lung cancer
...
population: NA
race: NA
sample type: tissue

**Fig. 4** GPT-4 Correction of BioSample metadata using a CEDAR template.

## Evaluation Metrics

To evaluate how our metadata-correction methods influenced searchability, we first needed to establish a ground truth against which to compare retrieval results. Because no formal gold standard was available, we manually examined the "tissue" values in our test dataset and developed simple rules to assign the correct tissue values (Table 2). For example, 'blood' is a commonly found tissue value, accounting for almost 50% of samples from BioSample (Figure 2). However, the metadata value for blood is often confounded by the methods of sample preparation, such as PBMC, whole blood, blood sample, plasma etc. Hence, we used a heuristic approach (Table 2) to assign the correct field value. This approach allowed us to construct an approximate gold standard that served as the best available reference for evaluating the performance of our methods.

We evaluated search performance on our two test datasets using standard information retrieval metrics—precision, recall, and F1-score (see Table 3 for definitions).

---

**Input:** Metadata record

**Output:** Assigned tissue label (lung, liver, ovary, blood, plasma, lymph, or unknown)

1. Initialize tissue field label as "unknown"
2. If tissue field contains the word "lung", set label to "lung"
3. Else if tissue field contains the word "liver" or "HCC", set label to "liver"
4. Else if tissue field contains the word "ovary" or "ovarian", set label to "ovary"
5. Else if tissue field contains the word "PBMC" or "blood", set label to "blood"
6. Else if tissue field contains the word "plasma", set label to "plasma"
7. Else if tissue field contains the word "lymph", set label to "lymph"
8. Return the assigned tissue label

**Table 2** Annotation rules used to construct our approximate gold standard dataset. These rules are used to assign corrected tissue values in the test dataset.

| Metric | Explanation | Formula |
|---|---|---|
| Precision | Precision is the ratio of correctly retrieved relevant instances to the total retrieved instances. | $\frac{\text{True Positives}}{\text{True Positives} + \text{False Positives}}$ |
| Recall | Recall is the ratio of correctly retrieved relevant instances to the total relevant instances. | $\frac{\text{True Positives}}{\text{True Positives} + \text{False Negatives}}$ |
| F1-Score | F1-Score is the harmonic mean of precision and recall. | $\frac{2 \times \text{Precision} \times \text{Recall}}{\text{Precision} + \text{Recall}}$ |

**Table 3** Definition of Search Metrics: Precision, Recall, and F1-Score

## Results

We found that recall improved substantially across all datasets and methods when we augmented GPT-4 with structured knowledge sources, compared to retrieval using the original baseline metadata. Figures 5-7 show the average recall, precision and F1-scores. In the BioSample dataset, the average recall increased from 20% with baseline metadata to 44% with GPT-4 guided by the data dictionary (GPT-4+DD), and further to 82% with GPT-4 guided by CEDAR templates (GPT-4+CEDAR). In GEO, recall improved from 15% at baseline to 32% with GPT-4+DD and to 44% with GPT-4+CEDAR. When we averaged results across both datasets, recall increased from 18% at baseline to 38% with GPT-4+DD and to 63% with GPT-4+CEDAR. We also observed an improvement in precision, rising from 58% at baseline to 66% with GPT-4+CEDAR (Figure 7). As a result, the overall F1-score increased from 24% to 63% (Figure 8).

We conducted statistical analyses using paired *t*-tests to compare recall values across conditions (baseline vs. GPT-4+DD, GPT-4+DD vs. GPT-4+CEDAR, and baseline vs. GPT-4+CEDAR). All comparisons showed statistically significant improvements ($p < 0.01$). We also computed effect sizes using Cohen's d, a standardized measure of the magnitude of differences between conditions. A Cohen's d value of 0.2 is typically considered a small effect, 0.5 a medium effect, and 0.8 or above a large effect. In our study, we observed large effect sizes ($d > 0.8$) for the improvements with GPT-4+CEDAR over baseline, indicating that the observed gains are not only statistically significant but also practically meaningful.

We observed that the most significant improvement in overall retrieval (F-1 score) and recall occurred when we standardized metadata using GPT-4 and CEDAR (GPT-4+CEDAR). The baseline metadata exhibited poor recall due to poor field value quality, which made it difficult to retrieve relevant records. By contrast, standardizing metadata with GPT-4 augmented with knowledge from CEDAR templates substantially improved retrieval accuracy and enhanced search performance.

Our qualitative analysis showed that declines in precision primarily stemmed from errors introduced by GPT-4 when it processed lengthy or ambiguous metadata. For example, GPT-4 occasionally changed correct tissue values—such as 'blood' to 'lung'—especially in longer-than-average BioSample records. Mentions of "lung" elsewhere in the metadata likely influenced these misclassifications. We also found that the rule-based labeling approach acted conservatively, often marking tissue values as unknown. In many such cases, GPT-4 correctly extracted the tissue value but was penalized due to inaccuracies in the gold standard.

In our dataset-specific analysis, we saw recall improvements across all queries. In the Liver Cancer dataset, searches for *tissue:liver* showed substantial improvements in recall, while precision dropped slightly due to a few minor misclassifications. For example, one of the retrieved metadata records was a match for the tissue type "liver," but our ground truth heuristic missed it because the relevant information appeared in the description field rather than the tissue field. In the *tissue:blood* query, we observed modest results where we either recorded the same recall values or a small improvement. Similarly, in the Ovarian Cancer dataset, the query for *tissue:ovary* led to substantial recall improvements with minor variation in precision. The Lung Cancer yielded the most promising results, with the *tissue:lung* query achieving high improvements in recall.

When we compared our two methods GPT-4+DD and GPT-4+CEDAR, we found that using only the repository's official data dictionary provided minor enhancements in recall for BioSample and significantly reduced performance for GEO. Since GEO records contain lengthy, free-text metadata values, there is increased potential for the LLM to deviate from the expected metadata values. For a non-small cell lung cancer (NSCLC) sample, GPT4+DD corrected the existing metadata to *tissue:NSCLC tumor* rather than the expected *tissue:lung*. In contrast, integrating CEDAR templates consistently resulted in noteworthy improvements across datasets, highlighting the importance of structured templates in the standardization of metadata.

We encountered difficulties when deploying alternative LLMs such as LLaMA-3, MedLLaMA-2, and Mistral. These models frequently produced formatting errors that required extensive post-processing, making them less appropriate for large-scale metadata standardization. In contrast, GPT-4+CEDAR achieved the best balance of accuracy and scalability, consistently outperforming other models in recall and retrieval performance.

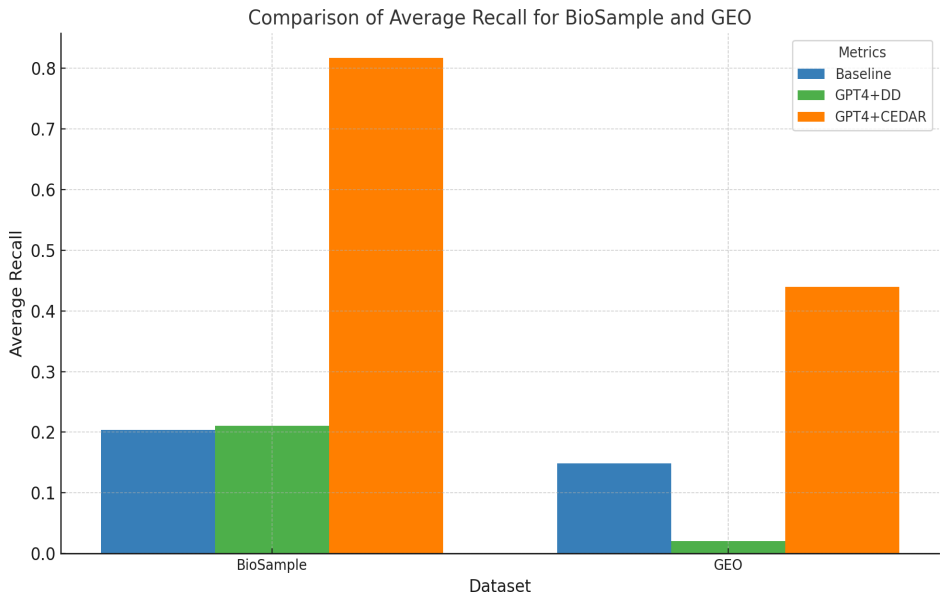

**Fig. 5** Comparison of average recall values for BioSample and GEO datasets across three result sets: Baseline, GPT-4+DD, and GPT-4+CEDAR.

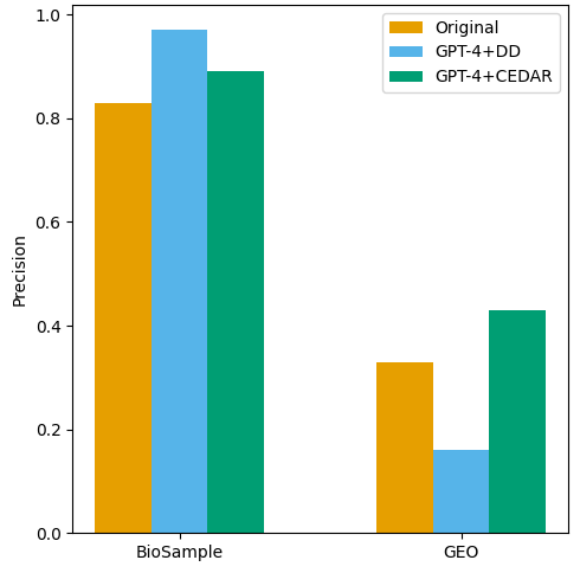

**Fig. 6** Comparison of average precision values for BioSample and GEO datasets across three result sets: Baseline, GPT-4+DD, and GPT-4+CEDAR.

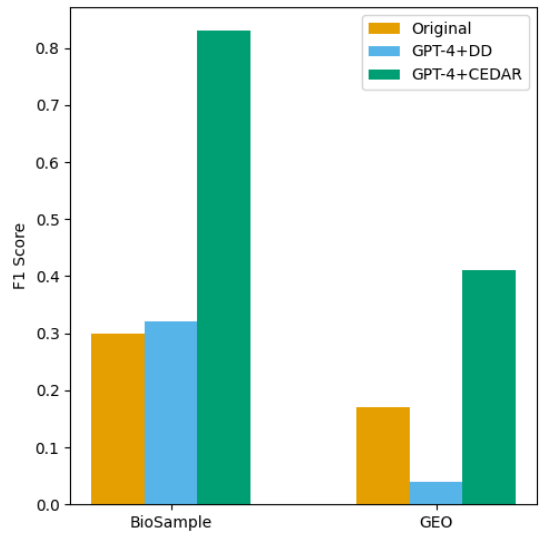

**Fig. 7** Comparison of average F1-scores for BioSample and GEO datasets across three result sets: Baseline, GPT-4+DD, and GPT-4+CEDAR.

**Discussion**

Our study empirically validates a long-standing but previously untested assumption at the heart of the FAIR data principles: that structured, standardized metadata improve data findability, accessibility, and reusability. While advocates of FAIR have promoted these principles for over a decade, most empirical work has focused on qualitative assessments or proxy metrics such as repository adherence or metadata availability. Our work directly quantifies the impact of metadata structure on dataset retrieval, providing clear evidence that aligning metadata with FAIR principles using structured templates significantly boosts performance in search systems. These findings fill a crucial evidence gap and offer a data-driven foundation for ongoing efforts to implement FAIR practices across the scientific data ecosystem.

Beyond this empirical contribution, our work addresses an emerging consensus in the field of AI: Achieving optimal performance with large language models requires more than just improved model architecture or increased scale. As LLMs become more powerful, their limitations—particularly in reasoning, precision, and domain specificity—have become more apparent. Increasingly, researchers are recognizing that enhancing LLM performance demands the strategic integration of structured, domain-specific, external knowledge [20, 21]. Our approach exemplifies this shift by embedding knowledge about metadata structures into LLM prompts through reusable, ontology-aligned metadata templates curated by experts. A range of methodologies has been proposed for injecting such knowledge into LLMs, including prompt engineering, fine-tuning with curated datasets, retrieval-augmented generation (RAG), integration with knowledge graphs, and neuro-symbolic approaches. Each of these methods offers a trade-off between accuracy, scalability, and annotation effort. We use *template-augmented prompt engineering* [22], a scalable and practical strategy for combining the linguistic fluency of LLMs with the semantic rigor of metadata templates.

Template-augmented prompt engineering occupies a unique space among knowledge infusion strategies. Fine-tuning, while effective, requires labor-intensive construction of labeled training data, which is often infeasible in specialized domains such as biomedicine where expert annotation is costly. Retrieval-augmented generation introduces scalability but increases architectural complexity and can cause inconsistencies between retrieved knowledge and generated outputs. By contrast, our approach requires no additional training or architectural modifications; instead, it leverages publicly available, community-developed metadata templates that encode expert-defined, ontology-aligned structures. These templates provide explicit, context-specific knowledge, guiding LLMs to generate accurate, standardized, and ontology-aligned metadata while reducing hallucinations and inconsistencies. Our method complements current knowledge-infusion strategies by providing a practical, reusable, and low-overhead strategy for augmenting LLMs with trusted domain knowledge, especially in high-stakes biomedical contexts where precision, reproducibility, and adherence to reporting guidelines and community standards (such as FAIR principles) are paramount.

Our findings further highlight the importance of manual review following automated metadata standardization. To support this work practice, we developed a tool that enables researchers to search using corrected metadata while simultaneously viewing the original, uncorrected metadata side-by-side. This interface helps users to compare both versions and aids informed decision-making about metadata trustworthiness, reducing cognitive load during secondary data analysis.

Our software is publicly available [23]. Such a tool is especially beneficial for secondary data users, who often need to assess metadata quality and compatibility across heterogeneous sources without detailed knowledge of the original study context.

Our findings are especially relevant in light of the growing interest in making scientific datasets "AI ready." Although this term is widely used, it often lacks operational definition. We propose that one core criterion for AI-readiness is the availability of structured, standardized, and machine-readable metadata that can reliably support automated data discovery, integration, and analysis. Scientific data repositories increasingly aim to serve as foundational infrastructure for AI-driven research, yet their metadata remains heterogeneous and inconsistently formatted. Our method offers a clear path toward remediation: by applying template-augmented LLMs to existing metadata records, repositories can enhance the semantic quality of their metadata, thus lowering the barrier to secondary analysis and enabling more robust AI applications. Our results thus offer clear, actionable insights for those seeking to prepare datasets explicitly for secondary AI-driven analyses, directly clarifying what operationalizing 'AI readiness' entails.

Secondary users of data—especially those conducting secondary analyses, exploratory studies, or hypothesis generation—stand to benefit the most from these improvements. These users often lack direct access to the original data collection context, and must rely entirely on metadata to interpret data correctly and to combine datasets from multiple sources. By automating the standardization of metadata into a form readily interpretable by computational tools, including AI systems, our approach significantly reduces preprocessing burdens and facilitates accurate computational analyses. Our empirical findings demonstrate that structured metadata dramatically improves dataset findability, accessibility, and reusability, offering concrete guidance for researchers and repository managers to ensure datasets meet established standards for data reuse. Thus, our work not only supports human-driven secondary analyses but also strengthens computational frameworks that increasingly underpin advanced AI-driven data exploration.

From a performance perspective, *recall* emerged as the most improved metric across our evaluations—a finding with direct implications for data discovery. In scientific data discovery, maximizing recall is critical: Relevant datasets may be rare, inconsistently described, or hidden behind poor-quality metadata. Hence secondary data analysts prioritize retrieving as many datasets as possible. Our results suggest that template-augmented prompt engineering can help to surface many datasets that would otherwise be missed when using legacy search systems. This improvement in findability is not merely technical; it has practical consequences for research reproducibility, comprehensiveness, and equity in access to scientific data.

Looking ahead, our work opens multiple pathways for future development. One promising direction is the integration of our method with retrieval-augmented generation (RAG) systems that can dynamically select the most relevant CEDAR template based on user-submitted metadata or search queries. Such a system would bring us closer to fully adaptive metadata correction pipelines that automatically tailor standardization strategies to the context of use. Additionally, we plan to scale our system to process the entire BioSample database—comprising approximately 5 million records—which would represent the largest effort to date in automated FAIR-aligned metadata remediation. The corrected database would not only improve discoverability and reuse for millions of samples but also serve as a valuable resource for benchmarking future metadata curation efforts.

Finally, our results speak to broader conversations in AI, knowledge representation, and scientific infrastructure. As LLMs become more integrated into scientific workflows, ensuring that they interact with structured, trustworthy, and domain-aware knowledge becomes paramount. Our work offers a concrete, reproducible model for doing so—one that is grounded in community standards, validated with empirical metrics, and designed with real-world applications in mind. Crucially, this study empirically confirms a foundational yet previously untested claim of FAIR advocates—that structured metadata significantly enhances dataset findability and reusability. This empirical validation offers advocates of FAIR principles robust, quantifiable proof of the effectiveness of metadata standardization—thus addressing a central critique that FAIR's impact had previously remained speculative. By demonstrating that FAIR-aligned metadata correction improves machine-mediated data access, we provide a missing link between principle and practice—and set the stage for a new generation of knowledge-aware, FAIR-compliant, AI-ready data ecosystems.

## Conclusion

Combining large language models with structured metadata templates is a transformative approach for improving the FAIRness of scientific data. Our findings demonstrate that domain-informed automation—specifically through CEDAR templates—enables substantial gains in metadata quality, directly enhancing data discoverability, accessibility, and reuse. We quantified these improvements through results in improved recall for biomedical datasets. This work shows that effective metadata standardization requires AI grounded in knowledge of metadata standards.

## Data Availability

The code and data are hosted at https://github.com/musen-lab/FAIRMetadataCuration

## Acknowledgments

This work was supported in part by grant R01 LM013498 from the U.S. National Library of Medicine.